\documentclass[twocolumn,aps,showpacs,longtable]{revtex4}

\usepackage{amsfonts,amsmath,graphicx,latexsym}
\usepackage{epsfig}
\usepackage{psfrag}
\newcommand{\vc}[1]{\mathbf{#1}}

\begin{document}
\title{Manifestation of a gap due to the exchange energy in a spinor condensate}
\author{Patrick Navez}
\affiliation{
Katholieke Universiteit Leuven,
Celestijnlaan 200 D,
Heverlee, Belgium \\ 
Universitaet Duisburg-Essen,
Lotharstrasse 1, 47057 Duisburg, 
Germany
}
%\begin{frontmatter}

%\date{\today}
\begin{abstract}
We investigate the dynamic response of population transfer between
two components of a finite temperature spinor Bose condensed gas
to a time-dependent coupling potential.
Comparison between results obtained in the Bogoliubov-Popov
approximation (BPA) and in the generalized 
random phase approximation (GRPA) shows noticeable 
discrepancies. In particular,
the inter-component current
response function calculated in the GRPA displays a gapped spectrum
due to the exchange interaction energy 
whereas the corresponding density response function is gapless. 
We argue that the GRPA is better since, contrary to the BPA,
it preserves the $SU(2)$ symmetry and the $f$-sum rule associated
to the spinor gas. In order to validate the approximation, we propose
an experimental setup that allows the observation of the
predicted gap.
\end{abstract}

%\begin{keyword}
%Spinor gas \sep  Bose-Einstein condensation \sep Gap in the excitation spectrum
%\end{keyword}
%\end{frontmatter}
\maketitle

\section{Introduction}

A great achievement in the discovery of condensation of Bose gas  
has been the experimental observation of the gapless and 
phonon-like nature of the collective excitations 
\cite{Ketterle, HP, Stringari}. The experimental measurement of 
the energy spectrum as a function of momentum has 
confirmed the validity 
%of the Hugenholtz-Pines theorem 
of the Bogoliubov theory \cite{books}. 
However, well understood at zero temperature, the finite 
temperature extension of this theory is still controversial. 
The main difficulty is to find an 
approximation both {\it conserving} and 
{\it gapless} \cite{HM}. 
The generalized random phase approximation (GRPA) is such an 
approach that has been introduced first  in the context of a 
quantum plasma in order to explain the collective phenomena \cite{Nozieres}. 
Its application 
to Bose condensation 
has been successful to determine  the sound velocity, 
the low-lying excitation spectrum 
observed
in a trapped Bose gas \cite{Reidl} and the critical 
velocity \cite{condenson}. Nevertheless, 
this theory  suffers from the apparent contradiction  
that its building blocks contain propagators
describing  the  particle
motion in terms of the gapped and parabolic 
Hartree-Fock (HF) mean field dispersion relation \cite{Reidl,Zhang}. 
This paradox is resolved  if 
the theory is reinterpreted adequately by making the distinction 
between collective excitations that result from oscillation 
of the condensate and quasi-particles or thermal excitations 
that compose  the normal part of the gas \cite{condenson}.
Generally, in other fluids the two kinds 
of excitation exist (e.g. the plasmons and the electrons in a plasma).
Therefore, no reason justifies that they should be identical 
in a Bose condensed gas as predicted from the Bogoliubov theory.

The only difference is that, in other fluids, the pole 
of the
one particle Green function  corresponds to the single particle spectrum
while the pole of the susceptibility function corresponds 
to the collective excitations.
In a Bose gas, below the condensation point, the two poles mix to
form new poles that are identical for the two functions \cite{HM,Reidl}.
In the non-conserving 
Bogoliubov theory, only one branch of excitation appears
as a pole which corresponds to the quasi-particle resulting
from hybridization of quantum states. 
In the GRPA, 
however, the two branches of the susceptibility
function and the branch of the one particle Green function mix
to form three common branches of excitations, 
one of which corresponds
to the phonon-like excitation \cite{Reidl}.     
In that case, it is not possible to associate a single
quasi-particle to the one particle Green function with
three branches of excitation as poles. For these reasons, the idea 
that the normal gas is composed of Bogoliubov-like quasi-particles seems 
too simplistic and must be reconsidered in the framework of  a 
conserving approach.  
Basically, the GRPA tells that the excitation spectrum
of a condensed
particle is phonon-like while the one for the thermal particle
is parabolic-like with a gap \cite{Zhang,condenson}. 
The whole difficulty is how to 
perturb the Bose gas in order to selectively address these excitations.

In this paper, we aim at verifying the validity of these hypothetical 
thermal excitations 
by proposing an experiment on a spinor Bose gas revealing the existence 
of the gap.
After reminding about the GRPA theory in section 2, 
%we start  
%to remind about the results about disturbance of a Bose gas subjected 
%to an external applied field for finite temperature. In the framework 
%of GRPA, we study more specifically the longitudinal and transverse 
%response functions. The first function disturb the whole gas and has 
%similar pole structure as the one particle Green functions and the 
%susceptibility functions namely the phonon-like excitations. The 
%second function, however, disturbs the normal part of the gas and appears 
%to have a completely different 
%a pole structure describing transition between thermal atoms 
%displaying the second kind of excitations. We compare this expression 
%with the one obtained in the Bogoliubov-Popov theory and 
%show how the dispersion 
%in the momentum does not allow easily to determine the nature of these 
%thermal excitations.
we study in section 3 the inter-component transition between two 
internal levels of the Bose gas subjected to an 
external coupling field \cite{Levitov,Fried,Sengstock,Cornell}.      
In particular, 
we are interested in determining the energy required 
to transfer
a thermal atom from a Bose condensed gas towards another 
internal sub-level. According to the GRPA, 
such an atom becomes distinguishable from 
the condensed atoms and consequently releases  
a fixed amount of  energy 
due to the Fock exchange interaction. An analog  
experiment is 
encountered in a superconductor-metal junction in which 
the measure of the electrical current as function of the voltage 
allows the determination of the BCS gap 
\cite{Feymann}. 
Both experimental \cite{Fried} and theoretical studies 
\cite{Levitov} have been carried out 
on transitions between two internal levels in order to 
probe the inter-component density response function. 
But 
an external coupling 
has the effect to rotate 
the whole gas in the spinor space at no energy cost 
in the long wavelength limit and thus this function is gapless. 
For this reason, we investigate 
instead the inter-component current response function in 
order to be able to address separately 
the thermal atoms from the condensed ones and to display the gap.
For comparison, a similar study is carried out in the 
gapless but non conserving Bogoliubov-Popov approximation 
(BPA) \cite{books,HM}. This approach extends the Bogoliubov 
theory to strongly depleted gas and leads 
to different results.
In section 4, we generalize the approach to the case of real 
experiments with 
a trap while section 5 ends with the conclusion.

\section{The GRPA formalism}

To start with, we consider a spinor Bose condensed gas of 
atoms with mass $m$ in a volume $\Omega$ populated only 
in two  
sub-levels chosen among the hyperfine structure of the atom 
and labeled by $\alpha=1,2$.
The scattering lengths between atoms of any sub-level 
are assumed to be equal to $a$ (in $^{87}Rb$ their relative 
difference is about $10^{-2}$ \cite{Sengstock}) and define 
the coupling constant $g=4\pi a/m$ ($\hbar=1$).
%The scattering length between atoms
%$a$ defines the coupling constant as
%$g=4\pi a/m$ ($\hbar=1$).
Defining the annihilation operators $c_{\alpha,\vc{k}}$ where 
$\vc{k}$ is the momentum and $\epsilon_\vc{k}=\vc{k}^2/2m$,
the Hamiltonian is written as:
\begin{eqnarray}
%\frac{H}{V}=\sum_{\alpha,\vc{k}} \frac{\vc{k}^2}{2m} 
%\rho^{\alpha,\alpha}_{\vc{k},\vc{0}}+
%\frac{g}{2}\sum{\vc{q}}
H=\sum_{\alpha,\vc{k}}\epsilon_\vc{k}c^\dagger_{\alpha,\vc{k}}
c^{}_{\alpha,\vc{k}}+ 
\frac{g}{2\Omega}
\!\!\!\!\! 
\sum_{\alpha,\beta,\vc{k},\vc{k'},\vc{q}}
%\sum_{\tiny \begin{array}{c}
%\vc{k},\vc{k'},\vc{q} \\
%\alpha,\beta
%\end{array}}
\!\!\!\!\!
c^\dagger_{\alpha,\vc{k}}c^\dagger_{\beta,\vc{k'}}
c^{}_{\alpha,\vc{k+q}}c^{}_{\beta,\vc{k'-q}}
\end{eqnarray} 
The eventual  energy shift between the two sub-levels 
is removed by choosing an appropriate 
representation of $c_{\alpha,\vc{k}}$.  
We shall study the dynamic evolution of
the excitation operator $\rho^{\alpha,\beta}_{\vc{k},\vc{q}}(t)=
e^{iHt}
c^\dagger_{\alpha,\vc{k}}c^{}_{\beta,\vc{k+q}}e^{-iHt}/\Omega$ 
in terms
of which we express the space Fourier transform of 
the densities and the currents, respectively 
$\rho^{\alpha,\beta}_{\vc{q}}(t)=
\sum_\vc{k} \rho^{\alpha,\beta}_{\vc{k},\vc{q}}(t)$ 
and $\vc{J}^{\alpha,\beta}_{\vc{q}}(t)=
\sum_\vc{k} \rho^{\alpha,\beta}_{\vc{k},\vc{q}}(t)
(\vc{k}+\vc{q}/2)/m$. These are associated to the 
$U(1)$ number conservation and $SU(2)$ spinor symmetries of $H$.
We first consider the case of a uniform condensed gas 
initially stable populated in the sub-level 1 such that 
$\langle \rho^{\alpha,\beta}_{\vc{k},\vc{q}} \rangle=
\delta_{\alpha,1}\delta_{\alpha,\beta}\delta_{\vc{q},\vc{0}}
n_{1,\vc{k}}$. $\langle \dots \rangle$ denotes 
the statistical ensemble average ${\rm Tr}(\sigma \dots)$
where $\sigma$ is the density matrix,    
$n_{1,\vc{k}}$ the momentum density in the mode $\vc{k}$ 
and $n_c= n_{1,\vc{0}}$ the condensate density.

Let us introduce a coupling between 
the two spinor components through the time 
dependent
Hamiltonian 
%:
%\begin{eqnarray}
$H_V(t)=
%\sum_{\vc{q}} 
\Omega
{\rho^{12}_{\vc{q}}}^\dagger
V^{12}_{\vc{q},\omega} e^{-i(\omega +i0) t}+c.c.
$
%\end{eqnarray}
where $V^{12}_{\vc{q},\omega}$ is   
a constant and $i0$ ensures the adiabatic condition. 
The change in the frequency mode $\omega$ of 
the inter-component 
density  
$\delta 
\rho^{12}_{\vc{q},\omega}$ from its 
equilibrium value $\langle \rho^{1,2}_{\vc{q}} \rangle=0$ induced 
by this potential is related  
to the inter-component susceptibility function through:
\begin{eqnarray}
\delta \rho^{12}_{\vc{q},\omega} 
%=\int_0^\infty
%dt \, e^{i(\omega+i0)t}  \delta \rho^{12}_{\vc{q}}(t)
=\chi_{12}(\vc{q},\omega)V^{12}_{\vc{q},\omega}
\end{eqnarray}
Formal linear response theory  allows to find the formula \cite{NP}:
\begin{eqnarray}\label{chi}
\chi_{12}(\vc{q},\omega)
=i \Omega \int_0^\infty \!\!dt\
e^{i(\omega+i0)t}\langle [{\rho^{12}_{\vc{q}}}^\dagger(0),
\rho^{12}_{\vc{q}}(t)]\rangle
\end{eqnarray}
%where $i0$ ensures integral convergence.
Although the initial state is not the $SU(2)$ symmetric equilibrium 
state, the linear response formalism is valid in a much more general 
context of a stable state such that $[H,\sigma]=0$.
Similarly to the intra-component case,  
provided that this function is causal and 
the $n_{\alpha,\vc{k}}$'s are isotropic, its imaginary part 
obeys the following f-sum rule \cite{Levitov}:
\begin{eqnarray}\label{sum}
\int_{-\infty}^\infty \!\!\!\! d\omega \, 
\omega \,{\rm Im}\chi_{12}(\vc{q},\omega)=
-\pi\frac{\vc{q}^2}{2m}n_1
%\begin{cases}
%\pi  
%(n_2-n_1) & p=0 \\
%-\pi\frac{\vc{q}^2}{2m}(n_1+n_2) & p=1 
%\end{cases}
\end{eqnarray}
This function is also connected to 
the longitudinal current response function 
$\chi^\parallel_{12}(\vc{q},\omega)$ through the exact relation:
\begin{eqnarray}\label{dJ}
\chi_{12}(\vc{q},\omega)=
\frac{n_1}{\omega+i0}+\frac{\vc{q}^2}{2m}
\frac{n_1}{(\omega+i0)^2}+
\frac{\vc{q}^2 \chi^\parallel_{12}(\vc{q},\omega)}{(\omega+i0)^2}
\end{eqnarray}
This last function is defined by replacing $\rho^{12}_{\vc{q}}(t)$
by $\vc{q}.\vc{J}^{12}_{\vc{q}}(t)/|\vc{q}|$ in (\ref{chi}).
%and obeys the sum rule:
%\begin{eqnarray}\label{sumJ}
%\int_{-\infty}^\infty \!\!\!\! d\omega \,
% {\chi^\parallel_{12}}''\!\!\!(\vc{q},\omega)=
%\frac{\pi}{m^2}\sum_\vc{k}\left(
%\frac{\vc{k}^2}{3}+\frac{\vc{q}^2}{4}\right)
%(n_{2,\vc{k}}-n_{1,\vc{k}})
%\end{eqnarray}
%All these sum rules 
The $f$-sum rule 
and Eq.(\ref{dJ}) result from the conservation
law $\partial_t \rho^{12}_{\vc{q}}(t)
=i\vc{q}.\vc{J}^{12}_{\vc{q}}(t)$ associated to the 
$SU(2)$ symmetry.

The GRPA approach is based on the assumption that 
the off-diagonal part of the excitation operator 
($\vc{q}\not=\vc{0}$) contributes less than the diagonal 
one ($\vc{q}=\vc{0}$) because the operators involved do not 
oscillate in phase but rather randomly \cite{condenson}. 
As a consequence, in the Heisenberg 
equation for $\rho^{12}_{\vc{k},\vc{q}}(t)$, we neglect 
all contributions quadratic in the off-diagonal operator 
in comparison with products involving off and  diagonal ones. 
In this way, we obtain an equation of motion of a similar 
form to the intra-component case \cite{condenson,Levitov}:
\begin{eqnarray}\label{op}
\left(i\partial_t -\epsilon_{2,\vc{k}+\vc{q}}^{HF}
+\epsilon_{1,\vc{k}}^{HF}\right)\rho^{12}_{\vc{k},\vc{q}}(t)
%\nonumber \\
=
%\left(n_{1,\vc{k}}-n_{2,\vc{k}+\vc{q}}\right)
gn_{1,\vc{k}}\rho^{12}_{\vc{q}}(t)
%+V^{12}_{\vc{q},\omega}e^{-i\omega t}/\Omega \right)
\end{eqnarray}
where $\epsilon_{\alpha,\vc{k}}^{HF}=\epsilon_\vc{k}
+gn_1(1+\delta_{1,\alpha})$ is the HF energy felt by the atom in 
sub-level $\alpha$.
%\sum_\beta n_\beta +gn_\alpha$.
%The only difference is that no integral 
%term due to exchange interaction appears between unlike atoms.  

\section{Results for the homogeneous gas}

The operatorial integral equation (\ref{op}) 
is solved exactly using methods 
developed in \cite{condenson}. Plugging the solution into 
(\ref{chi}), the susceptibility function
is found to be \cite{Zhang, Levitov}:
\begin{eqnarray}\label{chiRPA}
\chi_{12}(\vc{q},\omega)=\chi_{0,12}(\vc{q},\omega)/(1-g\chi_{0,12}(\vc{q},\omega))
\end{eqnarray}
where
\begin{eqnarray}\label{chi0}
\chi_{0,12}(\vc{q},\omega)= \sum_\vc{k}
\frac{n_{1,\vc{k}}}
%-n_{2,\vc{k+q}}}
{\omega +i0 + \epsilon^{HF}_{1,\vc{k}}-\epsilon^{HF}_{2,\vc{k+q}}}
\end{eqnarray}
As shown in \cite{Levitov}, this function is gapless 
since the pole $\omega =0$ corresponds to a rotation 
in the $SU(2)$ space.  
Using the  Kramers-Kronig relations \cite{books}, we  
check that Eq.(\ref{chiRPA}) fulfills 
the $f$-sum rule (\ref{sum}). Also, by introducing a vector 
potential as an external perturbation, the longitudinal current response 
function can be calculated as well in this approximation 
and the result verifies the property (\ref{dJ}).
Combination of (\ref{dJ}), (\ref{chiRPA}) and (\ref{chi0}) 
in the limit of long wavelength allows 
to obtain:
\begin{eqnarray}\label{long}
\chi_{12}^{\parallel}(\vc{q}=0,\omega)=
\frac{\sum_\vc{k} 2\epsilon_\vc{k}
n_{1,\vc{k}}/(3m)}{\omega +i0 +gn_1}
\end{eqnarray}
We find a non zero pole $-gn_1$ that corresponds 
to a gap 
due to the Fock exchange interaction and that is interpreted 
as the transition potential energy for  
any transferred thermal atom.
The fluctuations 
of the inter-component longitudinal current induce 
a relative motion between the normal fluid and the
superfluid in the spin space. 
Note that, contrary to the  intra-component case, the longitudinal
current response function Eq.(\ref{long}) does not have the same pole 
as the density response function Eq.(\ref{chiRPA}) and is 
identical to  the transverse current response function 
when $\vc{q} \rightarrow \vc{0}$.
Eq.(\ref{long}) is the 
main result of this paper and we shall show how this gap can be observed 
in experiments with a trap. 

In contrast, the use of the BPA leads to drastically  
different results. It 
corresponds to the 
Hartree-Fock-Bogoliubov approximation to the 
Hamiltonian in which the term quadratic in 
the product of anomalous average is also neglected \cite{books}. 
The chemical potential is 
given by $\mu=g(2n_1-n_c)$, the quantum operators describing 
the various  components evolve according 
to $c_{1,\vc{k}}(t)=e^{-i\mu t}(\sqrt{\Omega n_c}\delta_{\vc{k},0}
+u_{+,\vc{k}} 
e^{-i\epsilon^B_\vc{k}t} 
b_{1,\vc{k}}+ 
+ u_{-,\vc{k}} e^{i\epsilon^B_\vc{k}t}
b^\dagger_{1,-\vc{k}})$ and $c_{2,\vc{k}}(t)=
e^{-i(\mu +\epsilon_\vc{k}) t}
c_{2,\vc{k}}$.
$b_{1,\vc{k}}$ is the annihilation operator 
associated to the quasi-particle,  
$\epsilon_\vc{k}^B=\sqrt{2gn_c\epsilon_\vc{k} + 
\epsilon_\vc{k}^2}$ and 
$u_{\pm,\vc{k}}=\pm
((\epsilon_\vc{k}+gn_c)/2\epsilon^B_\vc{k}\pm 
1/2)^{1/2}$.
As a result, using (\ref{chi}), we obtain:
\begin{eqnarray}\label{BP}
\chi^{BP}_{12}(\vc{q},\omega)&=&
\frac{n_c}{\omega-\epsilon_\vc{q}+i0}+\sum_{\pm,\vc{k}}
\frac{u^2_{\pm,\vc{k}} (n^B_{1,\vc{k}}+\delta_{\pm,-})}{\omega+i0 \pm
\epsilon^B_{\vc{k}}-\epsilon^{}_{\vc{k}\pm\vc{q}}}
\\ \label{BPJ}
{\chi^{\parallel BP}_{12}}(\vc{0},\omega)&=&
\sum_{\pm,\vc{k}}\frac{2\epsilon_\vc{k}}{3m}
\frac{u^2_{\pm,\vc{k}} (n^B_{1,\vc{k}}+\delta_{\pm,-})}{\omega+i0 \pm
\epsilon^B_{\vc{k}}-\epsilon^{}_\vc{k}}
\end{eqnarray}
where $n^B_{1,\vc{k}}=\langle b^\dagger_{1,\vc{k}} 
b^{}_{1,\vc{k}}\rangle$.
Both approaches GRPA and BPA are gapless and their susceptibilities 
defined respectively by (\ref{chiRPA}) and (\ref{BP}) have a pole 
at $\omega=\epsilon_\vc{q}$ for zero temperature 
in agreement with the spectrum 
obtained from the linearization of the coupled Gross-Pitaevskii equations 
\cite{Fetter}. Besides this 
pole, we note from Eq.(\ref{BP}-\ref{BPJ}) that transitions 
in BPA result from scattering 
of a phonon into a normal atom releasing energy
$\epsilon^B_{\vc{k}}-\epsilon^{}_{\vc{k}+\vc{q}}$
and from creations of both a phonon and a normal atom 
absorbing the energy $\epsilon^B_{\vc{k}}+\epsilon^{}_{\vc{k}-\vc{q}}$.
As a consequence, for $\vc{q}=0$, the transition 
frequency spectrum is continuous with a line-width 
estimated to $gn_1$. 
In GRPA, however, 
only a process that releases the constant gap 
energy happens without any line-width. 

Therefore, the two theories display noticeable differences 
but strong theoretical arguments play in favor of GRPA. 
Due to the $SU(2)$ breaking of symmetry,  
the BPA violates the conservation laws and thus 
does not obey the $f$-sum rule 
nor  
Eq.(\ref{dJ}). In particular, for 
$\vc{q}=0$ and $\omega=0$, 
the action of the external coupling is to rotate 
the whole gas in the spin space while it rotates 
only the condensate in the BPA. Physically, this 
violation 
can be understood by noticing that
a phonon has not a well defined atom number 
($[b_{1,\vc{k}},\rho^{11}_{\vc{k},\vc{0}}]\not= b_{1,\vc{k}}$)
and therefore cannot be transformed into an atom 
alone during a transition. 
For these reasons, the GRPA is a better 
approximation that is justified if the predicted 
gap is observed in experiments. 

\section{Results for the trapped gas}

The considerations so far obtained 
are straightforwardly extended to a trapped gas.
We start with 
a condensed gas initially populated in the sub-level 1 confined 
in 
a parabolic potential
$V^{11}(\vc{r})=\sum_{i=x,y,z} \frac{1}{2} m\omega^2_i r_i^2$.
In this new situation, the density profile $n_1(\vc{r})$
depends 
on the position and is determined in good 
approximation 
from the mean field HF equations  
in the Thomas-Fermi limit  \cite{books}:
the condensate density $n_c(\vc{r})$ 
is determined from 
the
Gross-Pitaevskii equation in absence of the kinetic term 
$\mu=V^{11}(\vc{r})+g(2n_1(\vc{r})-n_c
(\vc{r}))$;
the thermal atoms density is determined 
from the semi-classical expression:
$n_{1,\vc{k}}(\vc{r})=
1/[\exp(\beta(\epsilon^{HF}_\vc{k}(\vc{r})+V^{11}(\vc{r})-\mu))-1]$ 
where $\beta=1/k_B T$ is the inverse temperature. 
Would we have applied the BPA, we should have rather used 
$n^B_{1,\vc{k}}(\vc{r})=1/(\exp(\beta \epsilon_\vc{k}^B(\vc{r}))-1)$ 
in the condensate region. 
We assume an isotropic trap 
with frequency
$\omega_i=\overline{\omega} =2\pi \times 100 Hz$, scattering length 
$a \sqrt{m \overline{\omega}}=10^{-3}$ and
$\mu=2\pi \times 2 kHz$. In the second sub-level, we consider 
both cases of a trapped and a flat potential 
$V^{22}(\vc{r})=V^{11}(\vc{r}),0$.
In order to induce a  
current in the trap and 
transfer atoms to the second sub-level, we 
apply the external coupling 
$V^{12}(\vc{r},t)=\vc{r_z}.\vc{F_z}e^{i\omega_0 t}$ 
during a finite time interval and beginning at $t=0$ 
where $\vc{F_z}$ refers to  a force constant along the z axis 
in the real space. 
%but rotating about the z axis in the spinor space.
This coupling can be produced either by  
a rotating magnetic field gradient 
$\vc{B}= B'r_z (\cos(\omega_0 t),\sin(\omega_0 t),0)$ that 
interacts with the gas with its spin orientation along the z-axis \cite{Sengstock}, or by 
a two photon Raman scattering between two multiplets (e.g. $F=1$ 
and $F=2$ in $^{87}Rb$) creating a stationary optical  
field 
with a node centered at the minimum of the trap \cite{Cornell}.

In presence of inhomogeneity, the density response 
has to be generalized into \cite{Stringari,books}:
\begin{eqnarray}\label{n12}
\delta \rho^{12}(\vc{r_1},t)=
\int_0^t \!\!\!dt'\!\!\! 
\int_\Omega \!\!\!d^3\vc{r_2}\,\chi_{12}(\vc{r_1},\vc{r_2},t-t')
V^{12}(\vc{r_2},t')
\end{eqnarray}
The total number of output atoms in the second level $N_2=
\int_\Omega  d^3 \vc{r}\, n_2(\vc{r},t)$ is 
determined through the transfer rate equation:
\begin{eqnarray}\label{N2}
\frac{dN_2}{dt}=i
\int_\Omega \!\!\!d^3\vc{r_1}\,
V^{21}(\vc{r_1},t) \delta \rho^{12}(\vc{r_1},t)+c.c.
\end{eqnarray}
For $t \gg 1/\omega_0$, the time integral can be extended to infinity in
(\ref{n12}) and Eq.(\ref{N2}) becomes time-independent. 
Combining Eqs.(\ref{n12}-\ref{N2}), rewriting $\vc{r_1}=\vc{r}+\vc{r'}/2$ and
$\vc{r_2}=\vc{r}-\vc{r'}/2$ and using
$\chi_{12}(\vc{r},\vc{q},\omega)=\int_0^\infty dt
\int_\Omega  d^3 \vc{r'}\, e^{-i(\vc{q}.\vc{r'}+\omega t)}
\chi_{12}(\vc{r_1},\vc{r_2},t)$, 
the transfer rate is expressed up to the second order in the coupling 
in terms
of the inhomogeneous susceptibility $\chi_{12}(\vc{r},\vc{q},\omega)$. 
This last function is calculated in the local density
approximation by replacing  the density and the frequency in the
homogeneous expressions (\ref{dJ})
and (\ref{long}) by their local forms $n_1(\vc{r})$ and
$\omega(\vc{r})=V^{11}(\vc{r})- V^{22}(\vc{r})-\omega_0$ \cite{Stringari}.
This procedure gives successively:
\begin{eqnarray}\label{res}
\frac{1}{F_z^2}\frac{dN_2}{dt}
&=& -2\pi \!\!\! \int d^3 \vc{r} \,
\left(r_z^2 + \frac{\partial^2_{q_z}}{4} \right)
{\rm Im} \left. \chi_{12}(\vc{r},\vc{q},\omega_0)
\right|_{\vc{q} \rightarrow 0}
\nonumber \\
&=&-2\pi \!\!\! \int d^3 \vc{r} \,
{\rm Im} \bigg[\frac{r_z^2 n_1(\vc{r})}{\omega(\vc{r}) +i0} 
+\frac{1}
{(\omega(\vc{r})+i0)^2}
\nonumber \\
&&\times \left(\frac{n_1(\vc{r})}{4m} + 
\frac{\sum_\vc{k} \epsilon_\vc{k} n_{1,\vc{k}}(\vc{r})/(3m)}
{\omega(\vc{r})+i0+gn_1(\vc{r})}\right) \bigg]
\end{eqnarray}
Eq.(\ref{res}) is expected to describe accurately the transfer 
rate provided $\omega_0 \gg \overline{\omega}$ and $t \ll 
\overline{\omega}^{-1}$ 
so that the effects 
of discretization in the excitation spectrum can be safely ignored.  
These effects also diminish due to the higher density states 
at finite temperature. 
Also, the linear response theory is valid provided that the typical 
time duration $t$ is much lower than the inverse of the typical Rabi 
frequency estimated to $F_z R_{TF}$ where $R_{TF}$ is the condensate size. 
This implies the upper bound 
$F_z \ll \overline{\omega}/R_{TF}\sim \overline{\omega}^2 \sqrt{m/2\mu}$. 
Thus, in GRPA, the total number of transferred atom results 
from a local rotation of the gas in the spin space 
in  regions where $\omega_0=V^{11}(\vc{r})- V^{22}(\vc{r})$ 
and 
from an inter-component current of thermal atoms releasing 
the exchange
Fock interaction energy in regions where
$\omega_0=gn_1(\vc{r})+V^{11}(\vc{r})- V^{22}(\vc{r})$.

\begin{figure}
%\psfrag{t}[c]{$T/T_c$}
%\psfrag{v}[b]{\rotatebox{270}{$v_s$}}
\begin{center}
\includegraphics*[width=7.0cm]{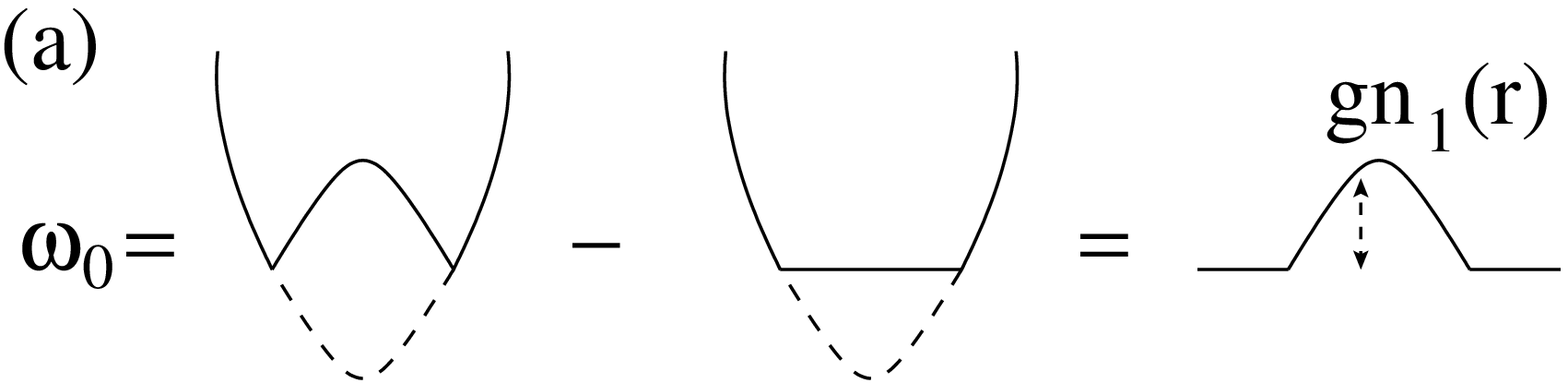}

\includegraphics*[width=7.0cm]{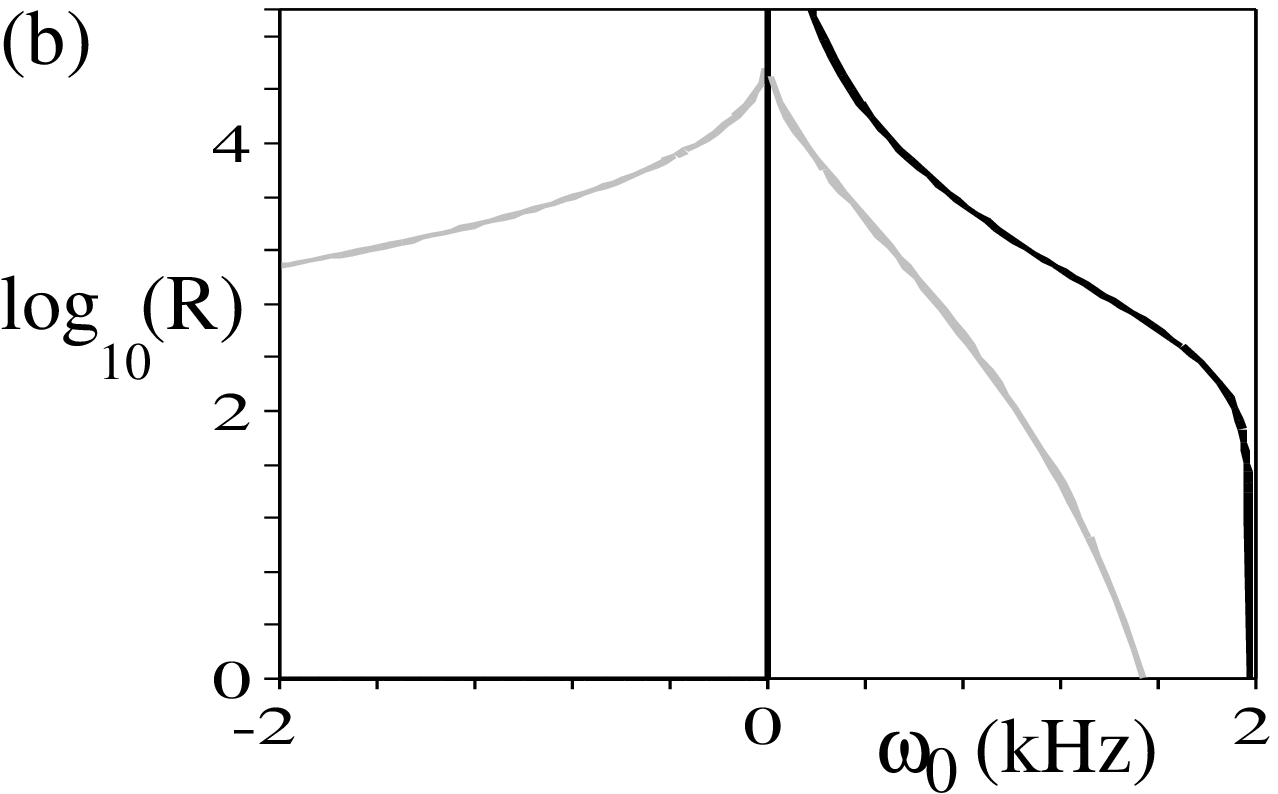}
\end{center}
\caption{In (a), 
schematic representation of the local potential energy (full line) 
in a trap felt by thermal atoms for the trapped sub-level 
1 (left) and 2 (right). The difference 
accounts for the transition energy. In (b), corresponding 
rate of output atoms in the second sub-level 
in GRPA (black) and in BPA (grey) for $T/T_c=0.56$.
}
\label{fig1}
\end{figure}

In the case of  
$V^{22}(\vc{r})=V^{11}(\vc{r})$, 
rotation in the spin space
occurs only for zero frequency so that a non zero
frequency rate provides a direct evidence of the gap 
$\omega_0=gn_1(\vc{r})$.
This released energy 
depends mainly on the local condensate density 
(see  Fig.\ref{fig1}a). Thus the transfer  
$R=m (\overline{\omega}/F_z )^2dN_2/dt$ calculated numerically 
from (\ref{res}) and shown in Fig.\ref{fig1}b 
as a function of  $\omega_0$ 
displays an inhomogeneous broadening.  
This rate decreases with frequency whereas the density 
of thermal atoms is higher in the edge of the condensate 
than in the center of the trap.
For comparison, we plot the transition 
rate obtained in BPA by inserting Eq.(\ref{BP}) in Eq.(\ref{res}) 
where contrary to GRPA energy absorption is possible when 
$\omega_0 < 0$. 
The reduced transition rate results from the broad 
spectrum displayed in  Eq.(\ref{BP}).

\begin{figure}
%[htb]
%\psfrag{t}[c]{$T/T_c$}
%\psfrag{v}[b]{\rotatebox{270}{$v_s$}}
\begin{center}
\includegraphics*[width=7.0cm]{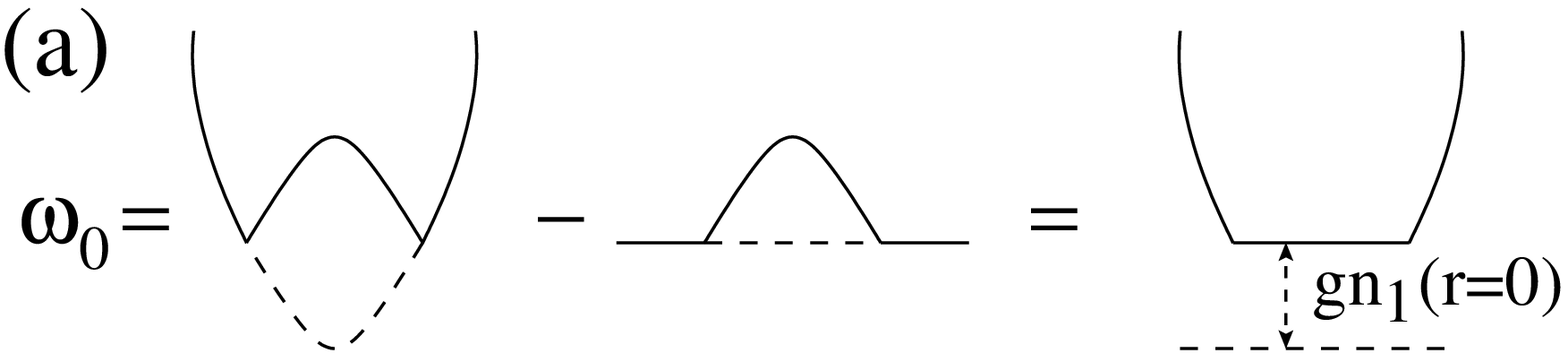}

\includegraphics*[width=7.0cm]{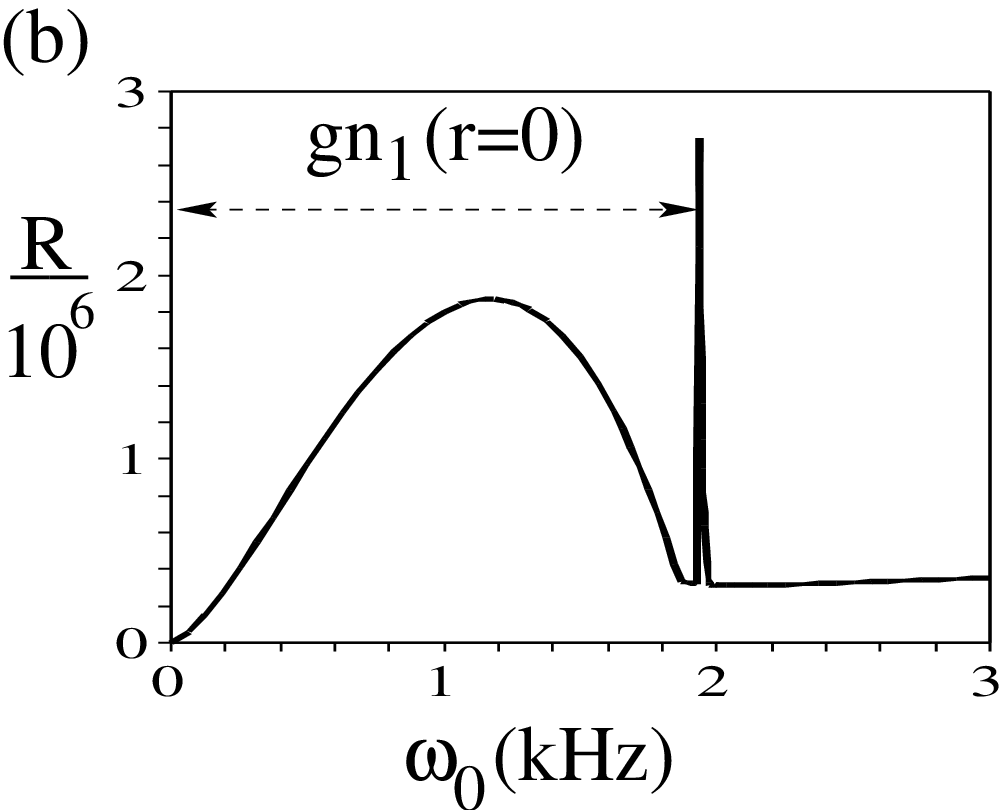}
\includegraphics*[width=7.0cm]{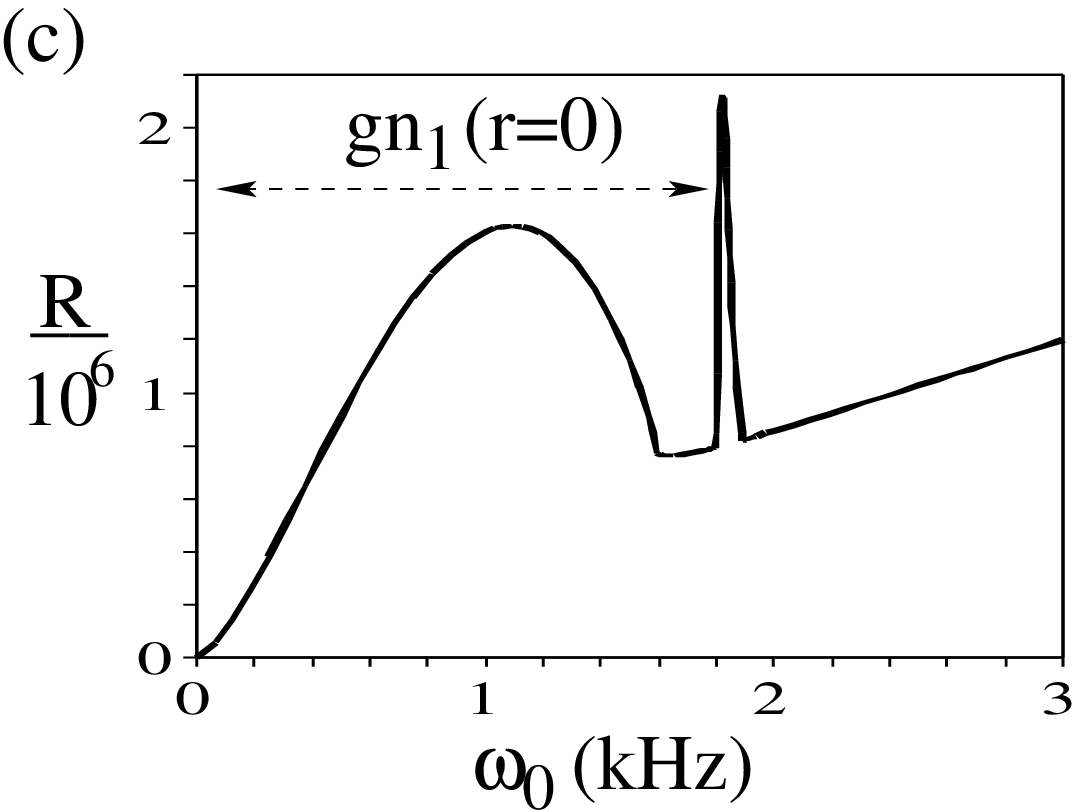}
\end{center}
\caption{In (a), 
same as Fig.1 but for a transition to an untrapped sublevel. 
Note the constant gap in the condensate region responsible for the peak.
Corresponding rate of output atoms 
for $T/T_c=0.56$ in (b) and $T/T_c=0.85$ in (c). }
\label{fig2}
\end{figure}

Fig.\ref{fig2} shows the same curves as Fig.\ref{fig1} but 
for the case $V^{22}(\vc{r})=0$.
The output atoms are not trapped in order 
to diminish the dispersion due to the inhomogeneity. As a result, 
a resonant transition occurs for a frequency close 
to $\omega_0=gn_1(\vc{r}=0)$. However the transition
due 
to the spinor rotation becomes frequency dependent and leads to a 
background noise. 
The broadening of the resonance is caused by 
the variation of the thermal density in the condensate 
region. The line-width is estimated to 
$\Delta \omega\sim g(mk_BT/2\pi)^{3/2}$ and increases 
with the temperature. Such a broader peak is observed in 
Fig.\ref{fig2}c  but at the expense of a more intense 
background noise. This noise is proportional to the local gas inertia 
$r_z^2 n_1(\vc{r})$ about the z-axis (see Eq.(\ref{res})) and is 
more important for stronger depletion in the region of high 
frequency. It considerably reduces if we choose 
$\omega_z \gg \overline{\omega}$ without altering the validity 
of the local density approximation 
($\omega_z \ll \omega_0$).

\section{Conclusions}

In conclusion, we analyze theoretically 
the atom transition rate between 
two sub-levels of a spinor condensate caused by a time dependent 
external coupling. Significant differences are reported 
between the GRPA and BPA models.
An experimental setup is suggested in order to validate 
at most one approach 
and can serve to understand the true nature of 
the quasi-particle dispersion relation.

%\centerline{\bf ACKNOWLEDGMENTS}
PN thanks K. Bongs for usefull discussions  and acknowledges support 
from the Belgian FWO project G.0115.06, from the 
Junior fellowship F/05/011 of the KUL research council,  
and from the German AvH foundation.

\end{document}